\documentclass[aip,jcp,reprint,superscriptaddress,longbibliography]{revtex4-2}%
\pdfoutput=1
\usepackage{graphicx}
\usepackage{color}
\usepackage{tabularx}
\usepackage{dcolumn}
\usepackage{longtable}
\usepackage{tensor}
\usepackage{placeins}
\usepackage{bm}
\usepackage{braket} 
\usepackage{amsmath}
\usepackage{amsfonts}
\usepackage{amssymb}
\usepackage{textcomp, gensymb}
\usepackage{float}%
\providecommand{\U}[1]{\protect\rule{.1in}{.1in}}

\newcommand{\eq}[1]{Eq.~(\ref{#1})} %
\def\be{\begin{equation}} %
\def\ee{\end{equation}} %
\newcommand{\bea}{\begin{eqnarray}}
\newcommand{\eea}{\end{eqnarray}}
\begin{document}
\title{Improving quantum measurements by introducing ``ghost" Pauli products}
\author{Seonghoon Choi}
\affiliation{Department of Physical and Environmental Sciences, University of Toronto Scarborough, Toronto, Ontario M1C 1A4, Canada}
\affiliation{Chemical Physics Theory Group, Department of Chemistry, University of Toronto, Toronto, Ontario M5S 3H6, Canada}
\author{Tzu-Ching Yen}
\affiliation{Chemical Physics Theory Group, Department of Chemistry, University of Toronto, Toronto, Ontario M5S 3H6, Canada}
\author{Artur F. Izmaylov}
\email{artur.izmaylov@utoronto.ca}
\affiliation{Department of Physical and Environmental Sciences, University of Toronto Scarborough, Toronto, Ontario M1C 1A4, Canada}
\affiliation{Chemical Physics Theory Group, Department of Chemistry, University of Toronto, Toronto, Ontario M5S 3H6, Canada}
\date{\today}
\begin{abstract}
Reducing the number of measurements required to estimate the expectation value of an observable is crucial for the variational quantum eigensolver to become competitive with state-of-the-art classical algorithms. To measure complicated observables such as a molecular electronic Hamiltonian, one of the common strategies is to partition the observable into linear combinations (fragments) of mutually commutative Pauli products. The total number of measurements for obtaining the expectation value is then proportional to the sum of variances of individual fragments. We propose a method that lowers individual fragment variances by modifying the fragments without changing the total observable expectation value.  Our approach is based on adding Pauli products (``ghosts”) that are compatible with members of multiple fragments. The total expectation value does not change because a sum of coefficients for each ``ghost'' Pauli product introduced to several fragments is zero. Yet, these additions change individual fragment variances because of the non-vanishing contributions of ``ghost” Pauli products within each fragment. The proposed algorithm minimizes individual fragment variances using a classically efficient approximation of the quantum wavefunction for variance estimations. Numerical tests on a few molecular electronic Hamiltonian expectation values show several-fold reductions in the number of measurements in the ``ghost” Pauli algorithm compared to those in the other recently developed techniques.
\end{abstract}
\maketitle
\graphicspath{{./Figures/}}

\section{\label{sec:intro}Introduction}

The variational quantum eigensolver (VQE)\cite{Peruzzo_OBrien:2014,McClean_Aspuru-Guzik:2016,Rybinkin_Izmaylov:2020,Cerezo_Coles:2021,Anand_Aspuru-Guzik:2022} is one of the most promising strategies for finding the ground-state of a molecular electronic Hamiltonian $\hat{H}$ on quantum devices in the noisy intermediate-scale quantum (NISQ) era.\cite{Preskill:2018} VQE circumvents the hardware limitations of NISQ devices by using both quantum and classical resources to find solutions of the time-independent Schr\"{o}dinger equation (TISE): $\hat{H} | \psi\rangle = E | \psi\rangle$. The lowest eigenvalue of the TISE can be found by minimizing the expectation value $E_{\theta} = \langle \psi_{\theta} | \hat{H} | \psi_{\theta} \rangle$ of $\hat{H}$ 
using a trial state, $|\psi_{\theta}\rangle$. In VQE, a quantum computer prepares $|\psi_{\theta}\rangle$ and measures the expectation value $E_{\theta}$, whereas a classical optimizer suggests a new trial state 
and minimizes $E_{\theta}$. 

The accurate measurement of $E_{\theta}$ is crucial for VQE and is not trivial because only Pauli-$\hat{z}$ operators can be measured on current digital quantum computers. The qubit Hamiltonian obtained by applying one of the fermion-qubit mappings\cite{Bravyi_Kitaev:2002, Seeley_Love:2012} to the molecular electronic Hamiltonian is a linear combination of $N_{q}$-qubit Pauli products:
\begin{equation}
\hat{H} = \sum_{k=1}^{N_{P}} c_{k} \hat{P}_{k}, \label{eq:H}
\end{equation}
where each Pauli product $\hat{P}_{k} = \otimes_{n=1}^{N_{q}} \hat{\sigma}_{n}$ is a tensor product of Pauli operators and identities for individual qubits: $\hat{\sigma}_{n} \in \{\hat{x}_{n}, \hat{y}_{n}, \hat{z}_{n}, \hat{1}_{n}\}$. It is, in principle, possible to obtain the expectation value $E_{\theta}$ by transforming $\hat{H}$ into its Ising form, 
$\hat{U} \hat{H} \hat{U}^{\dagger} = \sum_{i} a_{i} \hat{z}_{i} + \sum_{ij} a_{ij} \hat{z}_{i}\hat{z}_{j} + \dots$, and to measure using  
\begin{equation}
E_{\theta} = \langle \psi_{\theta} | \hat{U}^{\dagger} \hat{U} \hat{H} \hat{U}^{\dagger}  \hat{U}| \psi_{\theta} \rangle = \langle \hat{U} \psi_{\theta} | \hat{U} \hat{H} \hat{U}^{\dagger} | \hat{U} \psi_{\theta} \rangle.
\end{equation}
 However, this approach is impractical as it is equivalent to diagonalizing $\hat{H}$ in the computational basis. 

One of the most successful approaches for overcoming the difficulty in measuring $E_{\theta}$ is partitioning the Hamiltonian
\begin{equation}
\hat{H} = \sum_{\alpha=1}^{N_{f}} \hat{H}_{\alpha}
\end{equation}
and obtaining $E_{\theta}$ by adding every $\langle \psi_{\theta} | \hat{H}_{\alpha} | \psi_{\theta} \rangle$ measured separately.\cite{Izmaylov_Ryabinkin:2019, Jena_Mosca:2019, Zhao_Love:2020, Gokhale_Chong:2020, Verteletskyi_Izmaylov:2020, Yen_Izmaylov:2020, Shlosberg_Jena:2021, Crawford_Brierley:2021, Yen_Izmaylov:2021, Yen_Izmaylov:2022} These techniques partition the Hamiltonian such that it is feasible to implement on a quantum computer the unitary operator $\hat{U}_{\alpha}$ that transforms each $\hat{H}_{\alpha}$ into its Ising form. Such fragmentations of the molecular electronic Hamiltonian have been done in the qubit space\cite{Izmaylov_Ryabinkin:2019, Verteletskyi_Izmaylov:2020, Jena_Mosca:2019, Yen_Izmaylov:2020, Yen_Izmaylov:2022, Zhao_Love:2020, Gokhale_Chong:2020, Hamamura_Imamichi:2020} or in the fermionic space,\cite{Huggins_Babbush:2021, Yen_Izmaylov:2021} followed by the final transformations of the resulting operators into the qubit space. Fermionic algebra based techniques differ in whether they find the measurable fragments through the low-rank (LR)\cite{Berry_Babbush:2019, Motta_Chan:2021, Huggins_Babbush:2021} or full-rank (FR)\cite{Yen_Izmaylov:2021} decompositions of the two-electron part of the molecular electronic Hamiltonian in the second-quantized form.

On the other hand, the fully commuting (FC) fragmentation scheme,\cite{Yen_Izmaylov:2020, Crawford_Brierley:2021} as one of the qubit space techniques, partitions the qubit Hamiltonian into fragments containing mutually commutative Pauli products. Each unitary operator $\hat{U}_{\alpha}$ corresponding to an FC fragment can be implemented efficiently using only one- and two-qubit Clifford gates.\cite{Aaronson_Gottesman:2004,Yen_Izmaylov:2020, Crawford_Brierley:2021} If one were to impose a stricter condition of qubit-wise commutativity (QWC)\cite{Verteletskyi_Izmaylov:2020} between the Pauli products in a fragment, only one-qubit Clifford gates (i.e., local transformations) would be required to implement $\hat{U}_{\alpha}$. Yet, the performance of different fragmentation schemes should be judged based on the total number of measurements required to estimate $E_{\theta}$ up to desired accuracy $\epsilon$:\cite{Crawford_Brierley:2021, Yen_Izmaylov:2022}
\begin{equation}
M(\epsilon) = \frac{1}{\epsilon^{2}} \sum_{\alpha=1}^{N_{f}} \frac{\mathrm{Var}_{\psi}(\hat{H}_{\alpha})}{m_{\alpha}}, \label{eq:M_req}
\end{equation}
where $\mathrm{Var}_{\psi}(\hat{H}_{\alpha}) = \langle \psi| \hat{H}_{\alpha}^{2} |\psi \rangle - \langle \psi | \hat{H}_{\alpha} |\psi\rangle^{2}$ is the fragment variance, and $m_{\alpha}$ is the fraction of the total number of measurements allocated to the $\alpha$th independently measured fragment; from Eq.~(\ref{eq:M_req}) onward, we omit $\theta$ in $|\psi_{\theta}\rangle$ and $E_{\theta}$ for brevity. It was shown\cite{Yen_Izmaylov:2022} that $M(\epsilon)$ in the QWC techniques are typically higher than $M(\epsilon)$ in the FC techniques. Moreover, a recent study\cite{Bansingh_Izmaylov:2022} compared $M(\epsilon)$ in FC and QWC techniques while accounting for the non-unit fidelities of the quantum gates implementing $\hat{U}_{\alpha}$. This study demonstrated that even after considering fidelities of one- and two-qubit gates, the FC fragmentation methods usually have lower $M(\epsilon)$. As suggested by a recently performed resource estimation,\cite{Gonthier_Romero:2022} low $M(\epsilon)$ is crucial for VQE to be competitive with state-of-the-art classical methods. Therefore, we limit our discussion to the FC scheme.

Even within the FC approach, due to non-transitivity of the commutativity relation, there are multiple ways to partition the Hamiltonian to groups.\cite{Yen_Izmaylov:2020} Fragments' variances differ in different FC partitionings, and thus, $M(\epsilon)$ generally changes with a particular FC partitioning. 
In addition to changing the FC partitioning approach, one can lower $M(\epsilon)$ by exploiting that some Pauli products in the Hamiltonian can be shared by several measurable fragments: if $\hat{P}_{1}$ commutes with $\hat{P}_{2}$, and $\hat{P}_{2}$ commutes with $\hat{P}_{3}$, but $\hat{P}_{1}$ does not commute with $\hat{P}_{3}$ (e.g., $\hat{P}_{1} = \hat{x}_{1}, \hat{P}_{2} = \hat{x}_{1}\hat{x}_{2}, \hat{P}_{3} = \hat{z}_{1}\hat{z}_{2}$), then $\hat{P}_{1}$ and $\hat{P}_{3}$ must be placed in separate fragments $\hat{H}_{1}$ and $\hat{H}_{2}$, whereas $\hat{P}_{2}$ can be in both $\hat{H}_{1}$ and $\hat{H}_{2}$. Accordingly, Hamiltonian $\hat{H} = c_{1} \hat{P}_{1} + c_{2} \hat{P}_{2} + c_{3} \hat{P}_{3}$ can be fragmented into $\hat{H}_{1} = c_{1} \hat{P}_{1} + (c_{2} - a) \hat{P}_{2}$ and $\hat{H}_{2} = a \hat{P}_{2} + c_{3} \hat{P}_{3}$ while satisfying the sum rule $\hat{H} = \hat{H}_{1} + \hat{H}_{2}$ for any $a$. Then, $M(\epsilon)$ can be minimized with respect to $a$ if the corresponding 
variances for $\hat{H}_{1}$ and $\hat{H}_{2}$ are estimated. This reduction has been implemented in the iterative coefficient splitting (ICS) algorithm recently.\cite{Yen_Izmaylov:2022}  

A similar idea related to increasing the number of Pauli products that are measured in a single fragment has been employed in derandomization or biasing methods of classical shadow tomography.\cite{Hadfield_Mezzacapo:2022, Huang_Preskill:2021, Hilmich_Wille:2021, Wu_Yuan:2021, Hadfield:2021} In these techniques, the measurable fragments appear as a result of a particular unitary frame selection from the Clifford group for the Hamiltonian transformation. However, this choice of framework does not allow one to change the coefficients of 
Pauli products in measurable fragments and thus is less flexible than the ICS algorithm.\cite{Yen_Izmaylov:2022}

To reduce $M(\epsilon)$ even further, we explore an idea of more extensive fragment modification to lower their 
variances. The measurable fragments, $\hat H_\alpha$, do not need to be parts of the Hamiltonian; as long as
they sum to the Hamiltonian, the sum of their expectation values will be equal to $\bra{\psi}\hat H\ket{\psi}$.  
To lower variances of $\hat H_\alpha$ we will start with measurable fragments from 
the Hamiltonian and then introduce Pauli products that are not present in the Hamiltonian, ``ghost" Pauli products. 
The new ghost Pauli products $\hat{P}_{g}$ are compatible with several measurable fragments and are added to multiple measurable fragments such that the coefficients of $\hat{P}_{g}$ in those fragments cancel out upon summation (e.g., $\hat{H}_{1} \rightarrow \hat{H}_{1}^{\prime} \equiv \hat{H}_{1} + b \hat{P}_{g}$ and $\hat{H}_{2} \rightarrow \hat{H}_{2}^{\prime} \equiv \hat{H}_{2} - b \hat{P}_{g}$). 
Due to the cancellation, ghost Pauli products do not change the expectation value $E$. On the other hand, contributions of ghost Pauli products to $M(\epsilon)$ are non-vanishing, and one can reduce $M(\epsilon)$ by optimizing their coefficients. 

\section{Theory}
\label{sec:theory}

\subsection{Common framework for coefficient splitting and ghost Pauli products}
\label{subec:unified_framework}

It is convenient to introduce a unifying framework for coefficient splitting and ghost Pauli techniques by 
considering both techniques as modifications to the initial fragments $\hat{H}_{\alpha}^{(0)}$ 
obtained by partitioning the Hamiltonian: $\hat{H} = \sum_{\alpha} \hat{H}_{\alpha}^{(0)}$. 
Each of these initial fragments is a linear combination of mutually commutative Pauli products:
\begin{equation}
\hat{H}_{\alpha}^{(0)} = \sum_{k} c_{k} \hat{P}_{k}, \quad \hat{P}_{k} \in \mathcal{P}_{\alpha}^{(0)}, \quad \alpha = 1, \dots, N_{f}, \label{eq:no_groups}
\end{equation}
where $\mathcal{P}_{\alpha}^{(0)}$ denotes a set of Pauli products in the $\alpha^{\rm th}$ fragment, and $c_{k}$ are corresponding coefficients. 
In the coefficient splitting and ghost Pauli techniques, the initial fragments are modified into
\begin{equation}
\hat{H}_{\alpha} = \hat{H}_{\alpha}^{(0)} + c_{s}^{(\alpha)} \hat{P}_{s}, \quad \alpha \in \mathcal{A}_{s}
\end{equation}
by adding $c_{s}^{(\alpha)} \hat{P}_{s}$ to multiple fragments indexed by $\alpha \in \mathcal{A}_{s}$, where $\mathcal{A}_{s}$ corresponds to a set of fragments compatible with $\hat{P}_{s}$. In this formulation, 
the only difference between the coefficient splitting and ghost Pauli techniques is that for the former, 
the \emph{shared Pauli product} $\hat{P}_{s}$ is from $\hat{H}$, while this is not the case for the latter. 
The newly added coefficients $c_{s}^{(\alpha)}$ can be varied while ensuring $\hat{H} = \sum_{\alpha} \hat{H}_{\alpha}$ by imposing
\begin{equation}
\sum_{\alpha \in \mathcal{A}_{s}} c_{s}^{(\alpha)} = 0. \label{eq:sum_cs}
\end{equation}
Taking advantage of this increased freedom in the fragments, $M(\epsilon)$ is reduced by optimizing $c_{s}^{(\alpha)}$. In what follows, we will refer to the unified framework approach combining coefficient splitting 
and ghost Pauli products as the {\it shared Pauli products} (SPP) technique.

\subsection{Initial fragments}
\label{subsec:choosing_initial_frags}

The best choice of the initial set of fragments among the FC partitionings that gives the lowest $M(\epsilon)$ is the  sorted insertion (SI) technique.\cite{Crawford_Brierley:2021} It is based on the ``greedy" approach, and its success  can be understood by considering an $M(\epsilon)$ expression with the optimal choice of $m_\alpha$ minimizing 
$M(\epsilon)$\cite{Crawford_Brierley:2021} 
\begin{equation}
m_{\alpha} = \frac{[\mathrm{Var}_{\psi}(\hat{H}_{\alpha})]^{1/2}}{\sum_{j=1}^{N_{f}}[\mathrm{Var}_{\psi}(\hat{H}_{j})]^{1/2}}.\label{eq:m_alpha_opt}
\end{equation}
This choice of $m_\alpha$ gives
\begin{equation}
M_{\rm opt}(\epsilon) = \frac{1}{\epsilon^{2}} \left[\sum_{j=1}^{N_{f}} \sqrt{\mathrm{Var}_{\psi}(\hat{H}_{j})}\right]^{2}. \label{eq:M_opt}
\end{equation}
For distributions of fragments with a fixed sum of variances, the sum of square roots appearing in $M_{\rm opt}(\epsilon)$ is lower for fragments with an uneven distribution of variances. Such uneven 
distributions naturally appear in the ``greedy" approach. 

Even though the SI algorithm gives the best fragments for $M(\epsilon)$ without further optimization,  
to make the partitioning algorithm better for the SPP optimization, we used an additional modification. 
The modification serves the purpose of increasing the number of sharable Pauli products possible for 
created parts, thereby improving the performance of SPP. We modify the SI algorithm by avoiding grouping certain commuting 
Pauli products that otherwise would be grouped. Note that a product $\hat{P}_{k} \hat{P}_{l}$ of two 
Pauli products $\hat{P}_{k}$ and $\hat{P}_{l}$ is compatible with a measurable 
fragment $\hat{H}_{\alpha}^{(0)}$ that contains both these Pauli products (i.e., $\hat{P}_{k}, \hat{P}_{l} \in \mathcal{P}_{\alpha}^{(0)})$. More generally, all possible products $\hat{P}_{p} = \hat{X}_{1} \cdots \hat{X}_{L}$, where $\hat{X}_{k}$ is either $\hat{1}$ or $\hat{P}_{k} \in \mathcal{P}_{\alpha}^{(0)}$, commute with every $\hat{P}_{k} \in \mathcal{P}_{\alpha}^{(0)}$ (because every $\hat{P}_{k} \in \mathcal{P}_{\alpha}^{(0)}$ mutually commutes). 
In our SI modification, we avoid adding to an existing group any Pauli product that is a product of 
existing Pauli products (Appendix A describes details of this procedure). 

With this modification, our SI procedure has the following steps:

\textit{Initialization}: Sort the Pauli products $\hat{P}_{k}$ in the Hamiltonian 
$\hat{H}=\sum_{k} c_{k} \hat{P}_{k}$ in the descending order of $|c_{k}|$. To start, $N_{f} = 0$.

\textit{Iteration over the Pauli product $\hat{P}_{k}$ in the sorted list}:
\begin{enumerate}
\item If $N_{f} = 0$, go directly to step 2. Otherwise, iterate over $\alpha = 1, \dots, N_{f}$ and 
add $c_{k}\hat{P}_{k}$ to the first fragment $\hat{H}_{\alpha}^{(0)}$ 
if $\hat{P}_{k}$ commutes with all elements of  $\hat{H}_{\alpha}^{(0)}$ and is not a product 
of elements from $\hat{H}_{\alpha}^{(0)}$. 

\item If $c_{k} \hat{P}_{k}$ was not added to any fragment, increase $N_{f}$ by one and initiate a new fragment containing only $c_{k}\hat{P}_{k}$, i.e., $N_{f} \rightarrow N_{f} + 1$ and $\hat{H}_{N_{f}} = c_{k}\hat{P}_{k}$.
Go to step 1 unless every $\hat{P}_{k}$ in the Hamiltonian has already been considered.
\end{enumerate}

After this algorithm, a few of resulting fragments can be measured together 
because the Pauli products in these fragments are mutually commutative. Theorem~1 of Ref.~\onlinecite{Crawford_Brierley:2021} shows that merging such fragments 
always reduces $M(\epsilon)$. Therefore, we iterate through the pairs of fragments and merge them if all Pauli products in the two fragments are mutually commutative. 

\subsection{Selecting shared Pauli products}
\label{subsec:choosing_shared_paulis}

We select shared Pauli products for a pair of fragments at a time. It is possible to extend this approach to 
more than two fragments, but satisfying simultaneous compatibility for more fragments can reduce the number of 
possible sharable candidates. For a pair of fragments, Appendix A describes an efficient 
procedure to find all compatible Pauli products. The efficiency is due to reformulating the search procedure as a solution for a linear system of equations.
The search usually results in a large 
number of Pauli products, and only a fraction of these Pauli products can lower $M(\epsilon)$ appreciably. 
Thus, in what follows, we describe a two-step screening of Pauli products sharable between two fragments.

{\it 1. Removing redundancies:} Appendix B shows that if actions of two Pauli products ($\hat P_s$ and $\hat P_t$) 
on the system wavefunction are the same, $\hat{P}_{s} |\psi \rangle = \hat{P}_{t} | \psi \rangle$, 
there is no benefit in including them both as sharable Pauli 
products to lower $M(\epsilon)$. Using the involutory property of the Pauli products, $\hat P^2=1$, 
the action equality condition can be also rewritten as $\hat{P}_{t} \hat{P}_{s} |\psi \rangle = | \psi \rangle$. 
Since it is computationally difficult to test the equality condition for a general 
trial wavefunction $\ket{\psi}$, we use a Hartree--Fock approximation, $|\mathrm{HF}\rangle$, instead. 

$|{\mathrm{HF}}\rangle$ is an eigenstate of an all-$\hat{z}$ Pauli product, then 
$\hat{P}_{t} \hat{P}_{s} | \mathrm{HF} \rangle = |\mathrm{HF}\rangle$ is satisfied if $\hat{P}_{t} \hat{P}_{s}$ 
is an all-$\hat{z}$ Pauli product. Using that two Pauli products multiply to an all-$\hat{z}$ Pauli product if all the occurrences of either Pauli-$\hat{x}$ or Pauli-$\hat{y}$ operator are on the same qubits, we exclude all such Pauli products except one, e.g., only one from $\hat{x}_{1} \hat{y}_{2} \hat{z}_{3}$, $\hat{y}_{1} \hat{x}_{2}$, and $\hat{y}_{1} \hat{y}_{2}$ is included (see Appendix C for details). This preselection can lead to the removal of Pauli products that do not satisfy the action equality condition for $\ket{\psi}$ 
because $|\mathrm{HF}\rangle$ is only an approximation to $|\psi\rangle$. In particular, for multi-reference systems, the accuracy of the single-determinant $|\mathrm{HF}\rangle$ wavefunction decreases, and thus our redundancy removal procedure would also be less accurate. Therefore, we limit this removal procedure only to ghost Pauli products and not to the Pauli products in $\hat{H}$. 

{\it 2. Screening based on $M(\epsilon)$-lowering potential:}
The decrease in $M(\epsilon)$ resulting from the addition of $\hat{P}_{s}$ to a pair $\hat{H}_{\alpha}$ and $\hat{H}_{\beta}$ is
\begin{align}
\Delta M(\epsilon) &= \frac{1}{\epsilon^{2}} \sum_{j=\alpha,\beta} \frac{\mathrm{Var}_{\phi}(\hat{H}_{j}) - \mathrm{Var}_{\phi}(\hat{H}_{j}^{\prime})}{m_{j}} \nonumber \\
                   &= \frac{1}{\mu \epsilon^{2}} \left[2 c D(\hat{P}_{s}) - c^{2} \mathrm{Var}_{\phi}(\hat{P}_{s}) \right], \label{eq:delta_M}
\end{align}
where $\hat{H}^{\prime}_{\alpha} = \hat{H}_{\alpha} + c \hat{P}_{s}$ and $\hat{H}^{\prime}_{\beta} = \hat{H}_{\beta} - c \hat{P}_{s}$ are the fragments after the addition of $\hat{P}_{s}$; for brevity, 
\begin{equation}
D(\hat{P}_{s}) = \frac{m_{\alpha} \mathrm{Cov}_{\phi}(\hat{H}_{\beta}, \hat{P}_{s}) - m_{\beta} \mathrm{Cov}_{\phi}(\hat{H}_{\alpha}, \hat{P}_{s})}{m_{\alpha} + m_{\beta}}
\end{equation} 
and $\mu = m_{\alpha}m_{\beta}/(m_{\alpha} + m_{\beta})$ are defined. Note that we use the variances and covariances approximated using the configuration interaction singles and doubles (CISD) wavefunction $|\phi\rangle$ for selecting the shared Pauli products. If $\mathrm{Var}_{\phi}(\hat{P}_{s})=0$, from the Cauchy--Schwarz inequality
\begin{equation}
|\mathrm{Cov}_{\phi}(\hat{H}_{\alpha}, \hat{P}_{s})| \leq [\mathrm{Var}_{\phi}(\hat{H}_{\alpha})\mathrm{Var}_{\phi}(\hat{P}_{s})]^{1/2} \label{eq:cauchy_schwarz}
\end{equation}
for covariances, we find that $D(\hat{P}_{s}) = 0$; as a result, $\Delta M(\epsilon)$ in Eq.~(\ref{eq:delta_M}) also vanishes for all $c$. Therefore, we preselect only ghost Pauli products with variances $\mathrm{Var}_{\phi}(\hat{P}_{s}) > 0.9$. 

If $\mathrm{Var}_{\phi}(\hat{P}_{s}) \neq 0$, the reduction in $M(\epsilon)$ [Eq.~(\ref{eq:delta_M})] is maximized:
\begin{equation}
\Delta M_{\mathrm{max}}(\epsilon) = \frac{D(\hat{P}_{s})^{2}}{\mu \epsilon^{2}\mathrm{Var}_{\phi}(\hat{P}_{s})} \label{eq:M_max}
\end{equation}
when
\begin{equation}
c = \frac{D(\hat{P}_{s})}{\mathrm{Var}_{\phi}(\hat{P}_{s})}. \label{eq:opt_c}
\end{equation}
Using the Cauchy--Schwarz inequality~(\ref{eq:cauchy_schwarz}) and taking optimal $m_{\alpha}$ 
from \eq{eq:m_alpha_opt}, we find the upper bound for the expected reduction as
\begin{equation}
\Delta M_{\mathrm{max}}(\epsilon) \leq p \Lambda \equiv p \frac{[\mathrm{Var}_{\phi}(\hat{H}_{\alpha}) \mathrm{Var}_{\phi}(\hat{H}_{\beta})]^{1/2}}{[\mathrm{Var}_{\phi}(\hat{H}_{\alpha})]^{1/2} + [\mathrm{Var}_{\phi}(\hat{H}_{\beta})]^{1/2}}, \label{eq:M_max_max}
\end{equation}
where $\Lambda \leq \mathrm{min}[\mathrm{Var}_{\phi}(\hat{H}_{\alpha}), \mathrm{Var}_{\phi}(\hat{H}_{\beta})]^{1/2}$, and prefactor $p = 4 \sum_{j=1}^{N_{f}} [\mathrm{Var}_{\phi}(\hat{H}_{j})]^{1/2}/\epsilon^{2}$ which is common for all pairs. 
We preselect pairs of fragments based on Eq.~(\ref{eq:M_max_max}), which shows that large variances of the pair of fragments result in a large expected reduction in $M(\epsilon)$. After ranking the pairs according to Eq.~(\ref{eq:M_max_max}), we exclude all pairs in the first quartile. 

Among the preselected pairs of fragments ($\hat{H}_{\alpha}$, $\hat{H}_{\beta}$) and compatible Pauli products $\hat{P}_{s}$, we only add $\hat{P}_{s}$ to $\hat{H}_{\alpha}$ and $\hat{H}_{\beta}$ if this addition reduces $M(\epsilon)$ significantly [i.e., if $\epsilon^{2} \Delta M_{\mathrm{max}}(\epsilon) \geq 1 \cdot 10^{-5}$]. 
For each shared Pauli product $\hat{P}_{s}$, we iterate through the compatible pairs ($\hat{H}_{\alpha}, \hat{H}_{\beta}$) and evaluate the expected reduction in $M(\epsilon)$ [Eq.~(\ref{eq:M_max})] with measurements allocated according to the initial fragments: $m_{\alpha} \propto [\mathrm{Var}_{\phi}(\hat{H}_{\alpha}^{(0)})]^{1/2}$. If $\epsilon^{2}\Delta M_{\mathrm{max}}(\epsilon) \geq 1 \cdot 10^{-5}$, then $\hat{P}_{s}$ is added to $\hat{H}_{\alpha}$ and $\hat{H}_{\beta}$ with coefficients $c_{s}^{(\alpha)} = c$ and $c_{s}^{(\beta)} = -c$ [computed using Eq.~(\ref{eq:opt_c})]. The order in which $\hat{P}_{s}$ are added matters. The Pauli products compatible with a pair of fragments are not necessarily commutative with each other, and therefore the addition of $\hat{P}_{s}$ prevents all Pauli products that do not commute with $\hat{P}_{s}$ to be introduced. We iterate through the Pauli products in the descending order of the magnitude of their coefficients in the Hamiltonian ($|c_{k}|$). Therefore, all Pauli products in the Hamiltonian are introduced before the ghost Pauli products, while no particular order is imposed between the ghost Pauli products. The systematic analysis of how this ordering affects $M(\epsilon)$ is beyond the scope of our work.

\subsection{Optimization of coefficients and measurement allocation}
\label{subsec:ICSGS_optimization}

Once the fragments have been formed by introducing all selected shared Pauli products, the cost function $M(\epsilon)$ is optimized iteratively in two steps: 1) $m_{\alpha}$ are computed according to Eq.~(\ref{eq:m_alpha_opt}) with fixed $c_{s}^{(\alpha)}$, 2) $c_{s}^{(\alpha)}$ are optimized with fixed $m_{\alpha}$. For optimizing $c_{s}^{(\alpha)}$, we follow a similar procedure in ICS.\cite{Yen_Izmaylov:2022} To ensure that the fragments sum up to the Hamiltonian, we fix one of $\{c_{s}^{(\alpha)}\}_{\alpha \in \mathcal{A}_{s}}$ as $c_{s}^{(\ast_{s})} = - \sum_{\alpha \in \mathcal{A}_{s} \setminus \{\ast_{s}\}} c_{s}^{(\alpha)}$. The optimal coefficients are then found by solving the linear system
\begin{align}
&\epsilon^{2}\frac{\partial M(\epsilon)}{\partial c_{s}^{(\alpha)}} =\frac{2}{m_{\alpha}} \left[\mathrm{Cov}_{\phi}(\hat{H}_{\alpha}^{(0)}, \hat{P}_{s}) + \sum_{t:\alpha \in \mathcal{A}_{t}} \mathrm{Cov}_{\phi}(\hat{P}_{s}, \hat{P}_{t}) c_{t}^{(\alpha)} \right] \nonumber \\ &-\frac{2}{m_{\ast_{s}}} \left[\mathrm{Cov}_{\phi}(\hat{H}_{\ast_{s}}^{(0)}, \hat{P}_{s}) + \sum_{t:\ast_{s} \in \mathcal{A}_{t}} \mathrm{Cov}_{\phi}(\hat{P}_{s}, \hat{P}_{t}) c_{t}^{(\ast_{s})} \right] = 0. \label{eq:der_M_c}
\end{align}
The final measurement allocation $m_{\alpha}$ obtained at the end of this iterative procedure suggests optimal $M m_{\alpha}$ measurements for each fragment, where $M$ is the user-defined total number of measurements for estimating $\langle \psi| \hat{H} |\psi\rangle$. Even though $M m_{\alpha}$ is non-integer, because $M$ is $\sim 10^6$ in practice, rounding $M m_{\alpha}$ to the nearest integer should only have a negligible impact on the final error $\epsilon$. The algorithm proposed in this section will be referred as the shared Pauli products (SPP) method. 

As the number of introduced shared Pauli products increases, solving the linear system in Eq.~(\ref{eq:der_M_c}) can become difficult due to the cubic scaling with the number of optimization variables. Therefore, we propose an approximate optimization technique that scales linearly with the total number of Pauli products shared between pairs of fragments. For every selected $\hat{P}_{s}$ in a pair $\hat{H}_{\alpha}$ and $\hat{H}_{\beta}$, the optimal coefficients are assigned consecutively as $c_{s}^{(\alpha)} = c$ and $c_{s}^{(\beta)} = -c$, where $c$ is in Eq.~(\ref{eq:opt_c}). We will refer to the algorithm that employs this approximate optimization technique as sequentially optimized SPP (so-SPP). Other than the approach used for optimizing $c_{s}^{(\alpha)}$, SPP and so-SPP are identical in all other aspects.

In both SPP and so-SPP, the variances and covariances used for the optimization are only approximate, and thus $c_{s}^{(\alpha)}$ and $m_{\alpha}$ that minimize 
\begin{equation}
M_{\phi}(\epsilon) = \frac{1}{\epsilon^{2}} \sum_{\alpha=1}^{N_{f}} \frac{\mathrm{Var}_{\phi}(\hat{H}_{\alpha})}{m_{\alpha}}
\end{equation}
can still yield high values of $M(\epsilon)$ evaluated for the exact wavefunction. We refer to this problem as over-optimization because its origins are in accurate optimization based on approximate wavefunction. This over-optimization was observed for molecules near their equilibrium geometries in Ref.~\onlinecite{Yen_Izmaylov:2022} and is expected to be even more profound in strongly correlated systems for which the CISD wavefunction is less accurate. Appendix~D describes our approach to avoid over-optimization.

\section{\label{sec:num_dem}Results and Discussion}
We compare $M(\epsilon)$ in SPP with that in SI\cite{Crawford_Brierley:2021} and ICS.\cite{Yen_Izmaylov:2022} The SI algorithm and the ICS method were chosen because they have one of the lowest $M(\epsilon)$ among previously proposed fragmentation techniques. Following the study\cite{Yen_Izmaylov:2022} that originally proposed ICS, the initial fragments in the ICS method were obtained using the standard SI algorithm\cite{Crawford_Brierley:2021} because SI fragments already provide enough compatible Pauli products for coefficient splitting. The algorithms were used to compute $M(\epsilon)$ for electronic Hamiltonians of several molecules: H$_{2}$, LiH, BeH$_{2}$, H$_{2}$O, and NH$_{3}$. The qubit Hamiltonians were obtained using the Jordan--Wigner (JW) and Bravyi--Kitaev (BK)\cite{Bravyi_Kitaev:2002, Seeley_Love:2012} transformations of the fermionic Hamiltonians in the STO-3G basis set and the following nuclear geometries: $R(\mathrm{H} - \mathrm{H}) = 1\,$\AA (for H$_{2}$), $R(\mathrm{Li} - \mathrm{H}) = 1\,$\AA\ (for LiH), $R(\mathrm{Be} - \mathrm{H}) = 1\,$\AA\ with $\angle\mathrm{H}\mathrm{Be}\mathrm{H}=180\degree$ (for BeH$_{2}$), $R(\mathrm{O} - \mathrm{H}) = 1\,$\AA\ with $\angle\mathrm{H}\mathrm{O}\mathrm{H}=107.6\degree$ (for H$_{2}$O), and $R(\mathrm{N} - \mathrm{H}) = 1\,$\AA\ with $\angle\mathrm{H}\mathrm{N}\mathrm{H}=107\degree$ (for NH$_{3}$). To demonstrate the methods' performance in the strongly correlated regime, we computed $M(\epsilon)$ for BeH$_{2}$ and H$_{2}$O  at geometries stretched to more than twice their equilibrium bond length geometries, i.e., $R(\mathrm{Be} - \mathrm{H}) = 3.0\,$\AA\ and $R(\mathrm{O} - \mathrm{H}) = 2.2\,$\AA. Because Hamiltonian tapering\cite{Bravyi_Temme:2017, Setia_Whitfield:2020} is commonly employed in VQE to reduce the qubit requirements,\cite{Kirby_Love:2021, Maupin_Andrew:2021, Lang_Izmaylov:2021} we account for both untapered and tapered Hamiltonians and show that the proposed algorithm works well for both types of Hamiltonians. 

Table~\ref{table:main_results} presents $\epsilon^{2} M(\epsilon)$ for different algorithms. $\epsilon^{2} M(\epsilon)$ values are equivalent to the number of required measurements in millions to obtain $\langle \psi | \hat{H} | \psi \rangle$ with $10^{-3}$~a.u. accuracy. Owing to the increased flexibility in the fragments offered by coefficient splitting and ghost spawning, SPP has a lower $M(\epsilon)$ than both SI and ICS for most systems. The only exceptions from this trend are tapered JW Hamiltonian of H$_2$O at the stretched geometry and H$_{2}$, where no fragmentation method was able to obtain a lower $M(\epsilon)$ than that in SI. Excluding these cases, on average, SPP has $M(\epsilon)$ that is a factor of 5.3 lower than $M(\epsilon)$ in SI and a factor of 1.5 lower than $M(\epsilon)$ in ICS. The largest reductions in $M(\epsilon)$ compared to SI and ICS were by factors of 8.8 and 1.8, respectively. Moreover, the $M(\epsilon)$ values in SPP are also significantly lower than those in the fermionic decomposition methods. For systems presented both here and in Ref.~\onlinecite{Yen_Izmaylov:2021}, $M(\epsilon)$ in SPP is, on average, 19 times lower than $M(\epsilon)$ in the LR decompositions and 14 times lower than $M(\epsilon)$ in the FR decomposition. Neither the choice between the JW and BK transformations nor the qubit tapering produced an appreciable change in the number of measurements.

\begin{table}[h!]
\caption{Comparison of $\epsilon^{2} M(\epsilon)$ in SI,\cite{Crawford_Brierley:2021} ICS,\cite{Yen_Izmaylov:2022} SPP, and so-SPP for untapered and tapered Hamiltonians of H$_{2}$, LiH, BeH$_{2}$, H$_{2}$O, and NH$_{3}$ ($N_{q}$ is the number of qubits). \label{table:main_results}}
\begin{ruledtabular}
\begin{tabular}{lccccc}
Systems & $N_{q}$ & SI & ICS & SPP & so-SPP \\ 
\hline
\multicolumn{6}{c}{\it untapered Hamiltonians; Jordan--Wigner} \\
H$_{2}$               & 4  & 0.136&0.136 &0.136&0.136 \\
LiH                   & 12 & 0.882&0.240 &0.158&0.172 \\
BeH$_{2}$             & 14 & 1.11 &0.479 &0.370&0.413 \\
BeH$_{2}$ (stretched) & 14 & 2.11 &1.27  &0.928&1.47  \\
H$_{2}$O              & 14 & 7.59 &1.52  &1.33 &1.54  \\
H$_{2}$O (stretched)  & 14 & 3.96 &1.56  &1.05 &1.73  \\
NH$_{3}$              & 16 & 18.8 &3.34  &2.24 &2.90  \\
 \multicolumn{6}{c}{\it untapered Hamiltonians; Bravyi--Kitaev} \\
H$_{2}$               & 4  & 0.136&0.136 &0.136&0.136 \\
LiH                   & 12 & 0.882&0.240 &0.155&0.169 \\
BeH$_{2}$             & 14 & 1.11 &0.479 &0.360&0.406 \\
BeH$_{2}$ (stretched) & 14 & 2.11 &1.27  &0.937&1.61  \\
H$_{2}$O              & 14 & 7.59 &1.52  &1.34 &1.54  \\
H$_{2}$O (stretched)  & 14 & 3.96 &1.56  &1.03 &1.72  \\
NH$_{3}$              & 16 & 18.8 &3.34  &2.23 &2.89  \\
 \multicolumn{6}{c}{\it tapered Hamiltonians; Jordan--Wigner} \\
H$_{2}$               & 1  & 0.136&0.136 &0.136&0.136 \\
LiH                   & 8  & 0.860&0.240 &0.145&0.173 \\
BeH$_{2}$             & 9  & 1.42 &0.610 &0.341&0.388 \\
BeH$_{2}$ (stretched) & 9  & 2.27 &1.03  &0.945&1.31  \\
H$_{2}$O              & 10 & 7.38 &1.53  &0.952&1.33  \\
H$_{2}$O (stretched)  & 10 & 3.92 &1.01  &1.07 &3.45  \\
NH$_{3}$              & 14 &18.8  &3.33  &2.13 &2.84  \\
 \multicolumn{6}{c}{\it tapered Hamiltonians; Bravyi--Kitaev} \\
H$_{2}$               & 1  & 0.136&0.136 &0.136&0.136 \\
LiH                   & 8  & 0.885&0.217 &0.125&0.151 \\
BeH$_{2}$             & 9  & 1.43 &0.576 &0.322&0.374 \\
BeH$_{2}$ (stretched) & 9  & 2.23 &1.17  &0.847&1.82  \\
H$_{2}$O              & 10 & 7.48 &1.57  &0.916&1.31  \\
H$_{2}$O (stretched)  & 10 & 3.78 &1.22  &1.05 &2.73  \\
NH$_{3}$              & 14 & 18.8 &3.34  &2.23 &2.98  \\
\end{tabular}
\end{ruledtabular}
\end{table}

Increased flexibility of SPP leads to a lower $M(\epsilon)$, but also results in the increased number of optimization 
variables ($N_{s}$). This makes SPP more computationally expensive than ICS. 
While both ICS and SPP use the same approach to optimize the variables, 
for systems in Table~\ref{table:NpvsNsplit}, $N_{s}$ in SPP scales as $N_P^{1.5}$, whereas $N_{s}$ in ICS scales linearly with $N_P$, and on average, $N_{s}$ in SPP is 2.3 times higher than that in ICS. This larger number of optimization variables is problematic because the optimization step, which involves solving a system of linear equations, has $\sim N_s^3$ scaling. As a more efficient alternative, we proposed the so-SPP method that optimizes the coefficients approximately using a linearly scaling algorithm. This so-SPP method is particularly useful for larger systems because $N_{s}$ grows as $N_{P}^{1.5}$. 

Due to its approximate optimization, so-SPP has a higher $M(\epsilon)$ than SPP. Nevertheless, $M(\epsilon)$ in so-SPP is still much lower than that in SI. Furthermore, so-SPP also performed better than ICS for most systems except the strongly correlated ones. This under-performance of so-SPP suggests that correctly accounting for the couplings between $c_{s}^{(\alpha)}$ becomes more important for their optimization in strongly correlated systems.

\begin{table}[h!]
\caption{Comparison of the number of shared Pauli product coefficients ($N_{s}$) in ICS, SPP, and so-SPP for systems presented in Table~\ref{table:main_results}  ($N_{P}$ is the number of Pauli product terms in the Hamiltonian).  \label{table:NpvsNsplit}}
\begin{ruledtabular}
\begin{tabular}{lcccc}
Systems& $N_{q}$ & $N_{P}$ & ICS & SPP and so-SPP \\ 
\hline
\multicolumn{5}{c}{\it untapered Hamiltonians; Jordan--Wigner} \\
H$_{2}$               & 4  & 15   &   20 &   14 \\
LiH                   & 12 & 631  & 2096 & 2620 \\
BeH$_{2}$             & 14 & 666  & 1868 & 2949 \\
BeH$_{2}$ (stretched) & 14 & 666  & 2270 & 4050 \\
H$_{2}$O              & 14 & 1086 & 2988 & 7930 \\
H$_{2}$O (stretched)  & 14 & 1086 & 3388 & 7202 \\
NH$_{3}$              & 16 & 3609 & 9942 & 34875\\
 \multicolumn{5}{c}{\it untapered Hamiltonians; Bravyi--Kitaev} \\
H$_{2}$               & 4  & 15   &   20 &   14 \\
LiH                   & 12 & 631  & 2096 & 2718 \\
BeH$_{2}$             & 14 & 666  & 1868 & 3048 \\
BeH$_{2}$ (stretched) & 14 & 666  & 2270 & 4149 \\
H$_{2}$O              & 14 & 1086 & 2988 & 7987 \\
H$_{2}$O (stretched)  & 14 & 1086 & 3388 & 7761 \\
NH$_{3}$              & 16 & 3609 & 9942 & 35875\\
 \multicolumn{5}{c}{\it tapered Hamiltonians; Jordan--Wigner} \\
H$_{2}$               & 1  & 3    &    2 &    2 \\
LiH                   & 8  & 558  & 1226 & 2496 \\
BeH$_{2}$             & 9  & 596  & 1303 & 2840 \\
BeH$_{2}$ (stretched) & 9  & 596  & 1471 & 4224 \\
H$_{2}$O              & 10 & 1035 & 2508 & 9008 \\
H$_{2}$O (stretched)  & 10 & 1035 & 2717 & 9750 \\
NH$_{3}$              & 14 & 3609 & 9918 & 34089\\
 \multicolumn{5}{c}{\it tapered Hamiltonians; Bravyi--Kitaev} \\
H$_{2}$               & 1  & 3    &    2 &    2 \\
LiH                   & 8  & 558  & 1514 & 2584 \\
BeH$_{2}$             & 9  & 596  & 1315 & 2937 \\
BeH$_{2}$ (stretched) & 9  & 596  & 1448 & 4535 \\
H$_{2}$O              & 10 & 1035 & 2524 & 9046 \\
H$_{2}$O (stretched)  & 10 & 1035 & 2593 & 9775 \\
NH$_{3}$              & 14 & 3609 & 9942 & 34183\\
\end{tabular}
\end{ruledtabular}
\end{table}

\section{Conclusion}
\label{sec:conclusion}

This work generalizes the class of approaches that take advantage of Pauli products compatible with multiple measurable fragments to reduce the number of measurements. As an extension to the previous methods, which limit such Pauli products to those in the Hamiltonian, we allow adding to the measurable fragments the Pauli products that are not in the Hamiltonian. To preserve the expectation value of the Hamiltonian, these Pauli products have to be introduced such that their coefficients sum to zero. Yet, because the number of measurements is a non-linear function of the Pauli product coefficients, the contributions of the newly introduced Pauli products to the number of measurements are non-vanishing. By optimally introducing these Pauli products, the proposed SPP method was able to achieve the number of measurements several times lower than that in existing methods. 

This reduction in the number of measurements achieved by SPP came at the expense of an increased number of optimization variables. However, depending on the availability of classical resources, the number of optimization variables in the proposed algorithm can be suitably reduced by changing the user-defined criteria for selecting the shared Pauli products. In addition, we proposed computationally less expensive so-SPP, which uses an optimization technique that scales linearly with the total number of Pauli products shared by pairs of fragments.

Lastly, although our implementation of SPP relied on the covariances between Pauli products computed with the CISD wavefunction, one could improve this implementation by using the accumulated VQE measurement results to enhance the covariance estimates.\cite{Shlosberg_Jena:2021} A possible future direction is developing an algorithm that continuously adapts the SPP fragments and measurement allocation according to these enhanced covariance estimates.

\section*{Data Availability}
The data that support the findings of this study are available from the corresponding author upon request.

\section*{Code Availability}
Some part of the code that supports the findings of this study is available in the OpenFermion\cite{McClean_Babbush:2020} and PySCF\cite{Sun_Chan:2018} libraries. The rest of the code is available from the corresponding author upon request.

\section*{Acknowledgments}
A.F.I. is grateful to Ilya Ryabinkin for insightful discussions. S.C. acknowledges financial support from the Swiss National Science Foundation through the Postdoc Mobility Fellowship (Grant No. P500PN-206649). A.F.I. acknowledges financial support from the Google Quantum Research Program and Zapata Computing Inc. This research was enabled in part by support provided by Compute Ontario and Compute Canada.

\section*{Competing interests}
The authors declare that there are no competing interests.

\section*{Appendix A: Checking Pauli product factorization and finding compatible Pauli products} 

Two procedures described in this appendix are based on the isomorphism between Pauli products and 
a linear symplectic space of vectors over the binary field.\cite{Bravyi_Temme:2017, Yen_Izmaylov:2020} 
Thus, first, this isomorphism will be described and then 
problems of Pauli product factorization into existing Pauli products and of search for Pauli 
products that are compatible with two groups of commuting Pauli products will be addressed.    

\subsection*{Isomorphism between the space of the Pauli products and linear symplectic vector space 
over the binary field}

The isomorphism is built by corresponding each Pauli product $\hat{P}$ in the space of the $N_{q}$-qubit to 
vector $\vec{P}$ from a $2N_{q}$-dimensional linear symplectic vector space $\mathcal{F}$ over the binary field $Z_{2}$ as 
\begin{equation}
(\vec{P}_{n}, \vec{P}_{N_{q}+n}) = \begin{cases}
(0,1), & \text{if the $n$th qubit of $\hat{P}$ is $\hat{z}_{n}$}, \\
(1,0), & \text{if the $n$th qubit of $\hat{P}$ is $\hat{x}_{n}$}, \\
(1,1), & \text{if the $n$th qubit of $\hat{P}$ is $\hat{y}_{n}$}, \\
(0,0), & \text{if the $n$th qubit of $\hat{P}$ is $\hat{1}_{n}$},
\end{cases} \label{eq:PP2PV}
\end{equation}
for $n = 1, \dots, N_{q}$. For example, if $N_{q} = 4$ and $\hat{P} = \hat{z}_{2} \hat{y}_{3} \hat{x}_{4}$, then the corresponding Pauli vector is $\vec{P} = (0011;0110)$, where the semicolon only serves to improve the readability. 

Multiplication of Pauli products is related to addition of vectors: $\hat{P}_{1}\hat{P}_{2} = p\hat{P}$ is equivalent to $\vec{P}_{1} + \vec{P}_{2} = \vec{P}$, where $p$ is the phase factor unimportant for our application as it can be absorbed into the coefficient of $\hat{P}$ ($p = \pm 1, \pm i$ depending on the orders of single-qubit operators in $\hat{P}_{1} \hat{P}_{2}$). 

The commutativity of Pauli products is equivalent to the orthogonality 
of binary symplectic vectors, $\langle \vec{P}_{1}, \mathbf{J} \vec{P}_{2} \rangle = 0$,
where $\langle ., . \rangle$ is the usual Euclidean inner product, and $\mathbf{J}$ is 
the symplectic metric matrix
\begin{equation}
\mathbf{J} = \begin{pmatrix}
\bm{0}_{N_{q}} & \bm{1}_{N_{q}} \\
\bm{1}_{N_{q}} & \bm{0}_{N_{q}}
\end{pmatrix}
\end{equation}
consisting of $N_{q} \times N_{q}$ identity matrix ($\bm{1}_{N_{q}}$) and zero matrix ($\bm{0}_{N_{q}}$).

\subsection*{Checking factorization of Pauli products}
The mapping between $\hat{P}$ and $\vec{P}$ was exploited in the modified sorted insertion algorithm from Sec.~\ref{subsec:choosing_initial_frags}. In the algorithm, one of the two conditions for adding a Pauli product 
$\hat{P}$ to a group $\mathcal{P}_{\alpha}^{(0)}$ is that $\hat{P}$ is not a product of 
$\hat{P}_{k} \in \mathcal{P}_{\alpha}^{(0)}$. Because the multiplication of $\hat{P}$ corresponds to the addition of $\vec{P}$, a Pauli product $\hat{P}$ is not a product of $\hat{P}_{k} \in \mathcal{P}_{\alpha}^{(0)}$ if the corresponding Pauli vectors $\vec{P}$ and $\vec{P}_{k}$ ($k = 1, \dots, |\mathcal{P}_{\alpha}^{(0)}|$) are linearly independent. 
To test linear independence, we find the reduced row echelon form\cite{book_Press_Flannery:2007} of the matrix having $\vec{P}$ and $\vec{P}_{k}$ as rows. If there exists no all-zero row in this form, $\vec{P}$ and $\vec{P}_{k}$ are linearly independent, and therefore $\hat{P}$ is not a product of $\hat{P}_{k} \in \mathcal{P}_{\alpha}^{(0)}$.

\subsection*{Finding compatible Pauli products}
We also employed the mapping between $\hat{P}$ and $\vec{P}$ to find all Pauli products that are mutually commutative with every $\hat{P}_{k}$ in a group pair of $\hat{H}_{\alpha}$ and $\hat{H}_{\beta}$.  
When mapped to the vectors in $\mathcal{F}$, the mutual commutativity of $\hat{P}$ with 
every $\hat{P}_{k}$ is equivalent to 
\begin{equation}
\mathbf{M} \mathbf{J} \vec{P} = 0, \label{eq:kernel_eq}
\end{equation}
where $\mathbf{M}$ is a matrix having $\vec{P}_{k}$ as rows. Because all $\mathbf{J}\vec{P}$ satisfying Eq.~(\ref{eq:kernel_eq}) is in the null space of $\mathbf{M}$, we found the null space basis vectors using row reduction (Gaussian elimination).\cite{book_Press_Flannery:2007} Taking all possible linear combinations of these basis vectors yields every $\mathbf{J}\vec{P}$ satisfying Eq.~(\ref{eq:kernel_eq}). Every $\hat{P}$ compatible with $\hat{H}_{\alpha}$ and $\hat{H}_{\beta}$ is then obtained by applying $\mathbf{J}^{-1}$, which is equal to $\mathbf{J}$, on these $\mathbf{J}\vec{P}$ and mapping $\vec{P}$ back to $N_{q}$-qubit Pauli products using the inverse of Eq.~(\ref{eq:PP2PV}).

\section*{Appendix B: Identifying the redundant Pauli products}
We show that the required number of measurements, $M(\epsilon)$, cannot be reduced further by introducing $\hat{P}_{t}$ on top of $\hat{P}_{s}$ to a pair of measurable fragments ($\hat{H}_{\alpha}$ and $\hat{H}_{\beta}$) if 
\begin{equation}
\hat{P}_{s} | \psi \rangle = \hat{P}_{t} | \psi \rangle. \label{eq:redundant}
\end{equation}
Let us consider the reduction in $M(\epsilon)$ resulting from including both $\hat{P}_{s}$ and 
$\hat{P}_{t}$ as shared Pauli products between $\hat{H}_{\alpha}$ and $\hat{H}_{\beta}$:
\begin{equation}
\Delta M(\epsilon) = \frac{1}{\epsilon^{2}} \sum_{j=\alpha,\beta} \frac{\mathrm{Var}_{\psi}(\hat{H}_{j}) - \mathrm{Var}_{\psi}(\hat{H}_{j}^{\prime\prime})}{m_{j}}, \label{eq:M_diff}
\end{equation}
where $\hat{H}^{\prime\prime}_{\alpha} = \hat{H}_{\alpha} + c_{s} \hat{P}_{s} + c_{t} \hat{P}_{t}$ and $\hat{H}^{\prime\prime}_{\beta} = \hat{H}_{\beta} - c_{s} \hat{P}_{s}- c_{t} \hat{P}_{t}$ are the fragments after the addition of $\hat{P}_{s}$ and $\hat{P}_{t}$. The variances of the resulting fragments are
\begin{align}
\mathrm{Var}_{\psi}&(\hat{H}_{j}^{\prime \prime}) = \mathrm{Var}_{\psi}(\hat{H}_{j}) \nonumber \\ 
& + 2 c_{s} c_{t} \mathrm{Cov}_{\psi}(\hat{P}_{s}, \hat{P}_{t}) + c_{s}^{2} \mathrm{Var}_{\psi}(\hat{P}_{s}) + c_{t}^{2} \mathrm{Var}_{\psi}(\hat{P}_{t}) \nonumber\\
&\pm 2 [c_{s} \mathrm{Cov}_{\psi}(\hat{H}_{j}, \hat{P}_{s}) + c_{t} \mathrm{Cov}_{\psi}(\hat{H}_{j}, \hat{P}_{t})], \label{eq:Var_H_prime_prime}
\end{align}
where $\pm$ is $+$ for $j=\alpha$ and is $-$ for $j=\beta$. 

If Eq.~(\ref{eq:redundant}) holds, then $\mathrm{Cov}_{\psi}(\hat{H}_{j}, \hat{P}_{s}) = \mathrm{Cov}_{\psi}(\hat{H}_{j}, \hat{P}_{t})$ and $\mathrm{Cov}_{\psi}(\hat{P}_{s}, \hat{P}_{t}) = \mathrm{Var}_{\psi}(\hat{P}_{s}) = \mathrm{Var}_{\psi}(\hat{P}_{t})$ hold; therefore, Eq.~(\ref{eq:Var_H_prime_prime}) simplifies to
\begin{equation}
\mathrm{Var}_{\psi}(\hat{H}_{j}^{\prime \prime}) = \mathrm{Var}_{\psi}(\hat{H}_{j}) + \tilde{c}^{2} \mathrm{Var}_{\psi}(\hat{P}_{s}) \pm 2 \tilde{c} \mathrm{Cov}_{\psi}(\hat{H}_{j},\hat{P}_{s}), \label{eq:Var_H_prime_prime_simp}
\end{equation}
where $\tilde{c} = c_{s} + c_{t}$. Substituting Eq.~(\ref{eq:Var_H_prime_prime_simp}) in Eq.~(\ref{eq:M_diff}) yields
\begin{equation}
\Delta M(\epsilon) = \frac{1}{\mu \epsilon^{2}} \left[2 \tilde{c} D(\hat{P}_{s}) - \tilde{c}^{2} \mathrm{Var}_{\psi}(\hat{P}_{s}) \right], \label{eq:M_diff_simp} 
\end{equation}
where 
\begin{equation}
D(\hat{P}_{s}) = \frac{m_{\alpha} \mathrm{Cov}_{\psi}(\hat{H}_{\beta}, \hat{P}_{s}) - m_{\beta} \mathrm{Cov}_{\psi}(\hat{H}_{\alpha}, \hat{P}_{s})}{m_{\alpha} + m_{\beta}} 
\end{equation} 
and $\mu = m_{\alpha}m_{\beta}/(m_{\alpha} + m_{\beta})$ are defined the same as in Sec.~\ref{subsec:choosing_shared_paulis}. Equation~(\ref{eq:M_diff_simp}) shows that the maximum reduction in $M(\epsilon)$ obtained by optimizing $\tilde{c} = c_{s} + c_{t}$ is identical to that obtained by optimizing only $c_{s}$ without introducing $\hat{P}_{t}$ (i.e., with $c_{t}=0$), thereby demonstrating the redundancy of $\hat{P}_{t}$. [The resulting maximum reduction is given in Eq.~(\ref{eq:M_max}).]

\section*{Appendix C: Removing the redundant Pauli products}
In the proposed algorithm, the Pauli products satisfying Eq.~(\ref{eq:redundant}) are found using a Hartree--Fock approximation $|\mathrm{HF}\rangle$. As discussed in Sec.~\ref{subsec:choosing_shared_paulis}, if $|\mathrm{HF}\rangle$ is used, this action equality condition is equivalent to $\hat{P}_{t} \hat{P}_{s}$ product being an all-$\hat{z}$ Pauli product. The product of two Pauli products is an all-$\hat{z}$ Pauli product if all the occurrences of either Pauli-$\hat{x}$ or Pauli-$\hat{y}$ operator are on the same qubits. This condition can be checked conveniently by considering the first $N_{q}$ components of the corresponding $2N_{q}$-dimensional Pauli vectors ($\vec{P}_{n}$ for $n = 1, \dots, N_{q}$). Equation~(\ref{eq:PP2PV}) shows that $\vec{P}_{n}=1$ if $\hat{\sigma}_{n} = \hat{x}_{n}$ or $\hat{\sigma}_{n} = \hat{y}_{n}$ in $\hat{P} = \otimes^{N_{q}}_{n=1} \hat{\sigma}_{n}$; otherwise, $\vec{P}_{n}=0$. Therefore, $\hat{P}_{t} \hat{P}_{s}$ is an all-$\hat{z}$ Pauli product if the first $N_{q}$ components of $\vec{P}_{s}$ and $\vec{P}_{t}$ are identical. From now, we refer to Pauli products $\hat{P}_{s}$ and $\hat{P}_{t}$ as redundant if the first $N_{q}$-components of $\vec{P}_{s}$ and $\vec{P}_{t}$ are identical.

Appendix~B demonstrated that the addition of redundant Pauli products to a pair of fragments cannot reduce $M(\epsilon)$ further. Therefore, the proposed algorithm removes redundancies by excluding all but one element from each group of redundant Pauli products. For this, we must first group the Pauli products according to redundancy. Note that if $\hat{P}$ is redundant with one element of a group, then $\hat{P}$ is redundant with every element because redundancy is transitive (if $\hat{P}_{1}$ and $\hat{P}_{2}$ are redundant and $\hat{P}_{2}$ and $\hat{P}_{3}$ are redundant, then $\hat{P}_{1}$ and $\hat{P}_{3}$ are redundant). Therefore, $\hat{P}$ can be added to a group already if $\hat{P}$ is redundant with any one of the elements. Moreover, due to the transitivity, grouping the Pauli products according to redundancy is unique (unlike grouping them based on mutual commutativity). To group the Pauli products, we employed an algorithm similar to sorted insertion.\cite{Crawford_Brierley:2021} The algorithm for finding the redundancy groups has the following steps:

\textit{Initialization}: To start, $N_{g} = 0$, where $N_{g}$ is the number of redundancy groups. 

\textit{Iteration over Pauli products $\hat{P}_{s}$}:
\begin{enumerate}
\item If $N_{g} = 0$, go directly to step 2. Otherwise, iterate over all $N_{g}$ groups and add $\hat{P}_{s}$ to the group if $\hat{P}_{s}$ is redundant with the first Pauli product in the group.

\item If $\hat{P}_{s}$ was not added to any group, initiate a new group containing only $\hat{P}_{s}$ ($N_{g} \rightarrow N_{g} + 1$). Go to step 1 unless every $\hat{P}_{s}$ has already been considered.
\end{enumerate}

After finding the groups of redundant Pauli products, only one from each group is preselected as a potential sharable Pauli product. Because our redundancy condition is based on a Hartree--Fock approximation, the Pauli products in a redundancy group only satisfy the action equality condition, $\hat{P}_{s} |\psi\rangle = \hat{P}_{t} |\psi\rangle$, approximately. Therefore, there can still be small differences in the $M(\epsilon)$-lowering potential between the Pauli products in a group. In this work, we regard these differences as negligible and choose one Pauli product arbitrarily from each group.

\section*{Appendix D: Avoiding over-optimization}
While the aim of the optimization procedures in ICS, SPP, and so-SPP is to minimize $M(\epsilon)$, we can only minimize $M_{\phi}(\epsilon)$ instead because $\mathrm{Var}_{\psi}(\hat{P}_{k})$ and $\mathrm{Cov}_{\psi}(\hat{P}_{k}, \hat{P}_{l})$ evaluated with $|\psi\rangle$ are unavailable. To avoid over-optimization based on $M_{\phi}(\epsilon)$, we modify the cost function used in the optimization procedure (described in Sec.~\ref{subsec:ICSGS_optimization}) as 
\begin{equation}
\tilde{M}_{\phi}(\epsilon) = (1 - \delta) M_{\phi}(\epsilon) + \delta M_{\mathrm{avg}}(\epsilon),
\end{equation}
where $\delta = 0.001$, and
\begin{equation}
M_{\mathrm{avg}}(\epsilon) = \frac{1}{\epsilon^{2}} \sum_{\alpha=1}^{N_{f}} \frac{\mathrm{Var}_{\mathrm{avg}}(\hat{H}_{\alpha})}{m_{\alpha}}
\end{equation}
depends on the variances of $\hat{H}_{\alpha}$ averaged over all possible wavefunctions in the $2^{N_{q}}$-dimensional Hilbert space, i.e., $\mathrm{Var}_{\mathrm{avg}}(\hat{H}_{\alpha}) = \int \mathrm{Var}_{\psi} (\hat{H}_{\alpha}) d\psi $.
This average variance, in turn, depends on $\mathrm{Var}_{\mathrm{avg}}(\hat{P}_{k}) = 1 - \int \langle \psi | \hat{P}_{k} | \psi \rangle^{2} d\psi$ and $\mathrm{Cov}_{\mathrm{avg}}(\hat{P}_{k}, \hat{P}_{l}) = \int \mathrm{Cov}_{\psi}(\hat{P}_{k}, \hat{P}_{l}) d\psi$, which can both be evaluated analytically.\cite{Crawford_Brierley:2021} Using the solution\cite{Kalantre_Wang:2017} to exercise 7.3 of Ref.~\onlinecite{book_Watrous:2018}, one can obtain the integral in $\mathrm{Var}_{\mathrm{avg}}(\hat{P}_{k})$ as $I = \int \langle \psi | \hat{P}_{k} | \psi \rangle^{2} d\psi = 1/(1+2^{N_{q}})$, and thus find the average variance of any $\hat{P}_{k}$ as $\mathrm{Var}_{\mathrm{avg}}(\hat{P}_{k}) =  2^{N_{q}}/(1 + 2^{N_{q}})$. On the other hand, the average covariance between two non-identical Pauli products, $\mathrm{Cov}_{\mathrm{avg}}(\hat{P}_{k}, \hat{P}_{l})$, is zero according to Theorem 2 of Ref.~\onlinecite{Gokhale_Chong:2019}. 

Due to the change in the cost function from $M_{\phi}(\epsilon)$ to $\tilde{M}_{\phi}(\epsilon)$, the equations involved in the optimization [Eqs.~(\ref{eq:m_alpha_opt}), (\ref{eq:opt_c}), and (\ref{eq:der_M_c})] must also be modified. The modified equations are obtained by simply replacing every occurrence of $\mathrm{Var}_{\phi}(\hat{P}_{k})$ and $\mathrm{Cov}_{\phi}(\hat{P}_{k}, \hat{P}_{l})$ with  $\widetilde{\mathrm{Var}}_{\phi}(\hat{P}_{k}) = (1-\delta) \mathrm{Var}_{\phi}(\hat{P}_{k}) + \delta \mathrm{Var}_{\mathrm{avg}}(\hat{P}_{k})$ and $\widetilde{\mathrm{Cov}}_{\phi}(\hat{P}_{k}, \hat{P}_{l}) = (1-\delta)\mathrm{Cov}_{\phi}(\hat{P}_{k}, \hat{P}_{l}) + \delta \mathrm{Cov}_{\mathrm{avg}}(\hat{P}_{k}, \hat{P}_{l})$. Note that this modification is only done for the optimization step, and no such modification is performed for the equations involved in the shared Pauli product selection presented in Sec.~\ref{subsec:choosing_shared_paulis}.

\bibliography{ghost_pauli}

\end{document}